\documentclass{elsart}
\usepackage{graphicx}
\usepackage{dcolumn}
\usepackage{bm}
\begin{document}
\begin{frontmatter}
\title{Synchronization of delayed systems in the presence of delay time modulation} 
\author{Won-Ho Kye}, 
\ead{whkyes@empal.com}
\author{Muhan Choi},
\ead{ihshope@naver.com}
\author{Myung-Woon Kim},
\author{Soo-Young Lee},
\author{Sunghwan Rim}, and
\ead{rim@phys.paichai.ac.kr}
\author{Chil-Min Kim}
\ead{chmkim@mail.paichai.ac.kr}
\address{National Creative Research Initiative Center for Controlling Optical Chaos,
Pai-Chai University, Daejeon 302-735, Korea}
\author{Young-Jai Park}
\ead{yjpark@ccs.sogang.ac.kr}
\address{Department of Physics, Sogang University, Seoul 121-742, Korea}
\begin{abstract}
We investigate synchronization in the presence of delay time
modulation for application to communication.
We have observed that the robust synchronization is established by a common delay signal
and its threshold is presented using Lyapunov exponents analysis.
The influence of the delay time modulation in chaotic oscillators is also discussed. 
\end{abstract}
\end{frontmatter}

Synchronization phenomenon was discovered by Huygens in two coupled pendulum clocks\cite{Huy}.
The first description of synchronization in chaotic systems by Fujisaka and Yamada \cite{SyncOrg0}
gave new rise to scientific attention to the phenomenon 
and, over the past decade, 
synchronization in chaotic oscillators has been rigorously investigated 
not only for understanding of its role in nonlinear systems 
but for its potential applications in various fields \cite{SyncTut,SyncBook}. 
Synchronization in chaotic oscillators
is characterized by the fact that transverse variable ${\bf X}={\bf x}_1-{\bf x}_2$ converges 
to zero in the limit of $t \rightarrow \infty$, 
while longitudinal variable ${\bf x} = {\bf x}_1+{\bf x}_2$ is still in a chaotic state.  
Up to now the kinds of synchronization \cite{SyncTut,SyncBook,GPS} that 
have been observed include complete synchronization in identical systems
and phase synchronization (PS), lag synchronization,
and frequency synchronization  in slightly detuned systems,
while generalized synchronization  and 
generalized phase synchronization 
have been described in systems with different dynamics.

One of the most important fields where chaos has been practically applied is in communication.
The major concern in this field was that an encoded message is vulnerable to extraction
by  nonlinear dynamic forecasting \cite{NLDF}, 
when it is masked by the signal from the low-dimensional chaotic system.
Then it becomes essential to develop high-dimensional chaotic systems
having a multiple number of positive Lyapunov exponents
to implement a communication system. 
In this regard, one time-delayed system received a lot of attention:
that is, 
\begin{equation}
\dot{x}=f(x(t), x(t-\tau_0)),
\end{equation}
where $\tau_0$ is a delay time \cite{TD_Sync,TD_Laser,TD_Circuit,TD_Cont,TD_Comm}. 
The time-delayed system exhibits intriguing characteristics with increasing $\tau_0$.
Despite a small number of physical variables, 
the embedding dimension and number of positive Lyapunov exponents increases
as the delay time increases,
and the system eventually transits to hyperchaos\cite{Hyperchaos}. 
For this reason, a time-delayed system was considered to be the most suitable
candidate for implementing communication systems.

However, it was newly discussed the communication based on a time-delayed system 
is still vulnerable because the delay time $\tau_0$ can be exposed 
by several measures, e.g., autocorrelation \cite{AutoCorr}, filling factor\cite{Bunner}, 
and one step prediction error, etc \cite{Bunner,TD_ATT}. 
If the delay time is known, the time-delayed system collapses
to the simple manifold 
in the $(\dot{x}, x(t), x(t-\tau_0))$ space \cite{Bunner,TD_ATT}.
So the time-delayed system becomes quite a simple one to someone who knows the delay time $\tau_0$, 
and that message encoded by the chaotic signal
can be extracted by the common attack methods \cite{NLDF}.

Accordingly, the most crucial questions to address regarding to the application
of time-delayed system to communications come to be 
{\it "Is there any scheme which hides the delay time from an eavesdropper"}, 
and {\it "Is it possible to find out a robust synchronization regime under the scheme"}. 
In this paper, we consider a generalization of time-delayed systems in which 
the delay time is not constant but modulated in time, which was introduced
to study the general property of the delay differential system \cite{Madruga}.
Applying the generalization to two coupled chaotic systems, 
we focus our discussion on answering the two questions above.
Thus, we would show that there exists 
a robust synchronization regime even if the delay time is modulated and 
that the imprints of delay time in time series are completely
wiped out by the modulation of delay time. 

\begin{figure}
\begin{center}
\rotatebox[origin=c]{0}{\includegraphics[width=8.3cm]{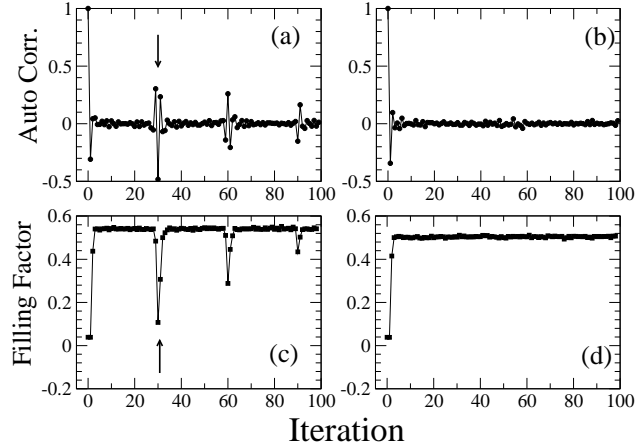}}
\caption{Autocorrelation functions and filling factors at $\lambda=4.0, \beta=0.7$:
(a) and (c) constant delay time $\tau_0=30$;
(b) and (d) delay time modulation with $g(t, \tau_0)= (0.95\tau_0/2)\sin(t)+\tau_0/2$ and $\tau_0=60$. 
The primary peaks in (a) and (c)
show the constant delay time $\tau_0$.}
\end{center}
\end{figure}
First, to explain the delay time modulation (DTM), 
we consider a single logistic map with time-delayed feedback,
and observe what an effect the modulated delay time $\tau$ 
has on the characteristics of the time series: 
$x(t+1)=\lambda \tilde{x}(t)(1-\tilde{x}(t))$,
where $\tilde{x}(t)= (1-\beta) x(t) +\beta x(t-\tau)$ with 
$\tau(t)=g(t, \tau_0)$.
The modulation function of the delay time is given by:
\begin{equation}
\tau=g(t, \tau_0).
\end{equation}
If we take $g(t,\tau_0)$ as a constant, 
the system returns to the usual time-delayed system.
As we shall see, the nontrivial choice of the modulation function
not only has a crucial effect on the characteristics of synchronization, 
but also remarkably transforms the characteristics of the time series.  
Figure 1 shows autocorrelations and filling factors \cite{FF} 
for two different modulations. 
In the case of the constant delay time (Fig. 1 (a) and (c)), 
one can easily notice that the peaks indicated 
by arrows exactly correspond to the multiples of the delay time $\tau_0$.
On the contrary, when the modulation is 
sinusoidal (Fig. 1(b) and (d)) the imprints of delay time are completely 
wiped out, and the detection of delay time seems 
to be impossible. 
We emphasize that the disappearance of the
imprints of delay time is caused by the fact that the delay time is
changing in time.
Accordingly, Figure 1 clearly shows that the results 
do not depend on the method of the detection.


To analyze synchronous behaviors, 
we consider a system of two coupled Lorenz oscillators 
with common time-delayed feedback:

\begin{equation}
\left .
\begin{array}{lll}
\dot{x}_1&=& \sigma (y_1-\tilde{x}_1), \\
\dot{y}_1&=& -\tilde{x}_1z_1 + r\tilde{x}_1-y_1, \\
\dot{z}_1&=& \tilde{x}_1 y_1 - b z_1, \\ 
	& & \\
{\tau} &=& g(t, \tau_0), 
\end{array}
\right \} \mbox{Transmitter} 
\end{equation}
\begin{equation}
\left .
\begin{array}{lll}
\dot{x}_2&=& \sigma (y_2-\tilde{x}_2),  \\
\dot{y}_2&=& -\tilde{x}_2 z_2 + r\tilde{x}_2-y_2, \\
\dot{z}_2&=& \tilde{x}_2 y_2 - b z_2,
\end{array}
\right \} \mbox{Receiver~~~}
\end{equation}
where $\sigma=10.0$, $r=28.0$, and $b=8/3$ and 
$\tilde{x}_1=(1-\beta) x_1(t) + \beta y_1(t-\tau)$ 
and  $\tilde{x}_2=(1-\beta) x_2(t) + \beta y_1(t-\tau)$.
Here we take the modulation function as 
$g(t, \tau_0)= \frac{0.95\tau_0}{2} \sin(\omega t) + \frac{\tau_0}{2}$ and $\tau_0=60$.
We note that the systems are coupled by common signal $\beta y_1(t-\tau)$. 
Figure 2 shows the trajectories in the phase space with
time-delayed feedback. One can see that the shape of the attractor
is so seriously deformed that it is impossible to identify 
the original shape of the Lorenz attractor.
This deformation of the attractor is the typical characteristic of 
the time-delayed system which stems from the transition 
to hyperchaos by time-delayed feedbacks \cite{TD_Sync,Bunner}.

\begin{figure}
\begin{center}
\rotatebox[origin=c]{0}{\includegraphics[width=8.3cm]{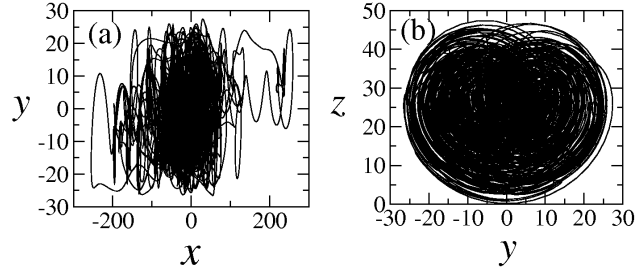}}
\caption{Projective trajectories of Eq. (3) when $\beta=0.93$ and $\omega=0.005$. 
	(a) in $(x,y)$ space (b) in $(y,z)$ space.}
\end{center}
\end{figure}

The synchronization transition can be easily detected by observing the
relative motion of the two oscillators. 
The equation of motion describing the relative 
motion is obtained from Eqs. (3)-(4):
\begin{eqnarray}
\dot{X} &=& \sigma (Y - (1-\beta) X), \nonumber \\ 
	& & \nonumber\\
\dot{Y} &=& -(1-\beta) (x_1Z +X z_1- X Z) \nonumber\\
	& &-\beta y_1(t-\tau) Z -r (1-\beta)X -Y, \\ 
	& & \nonumber\\
\dot{Z} &=& (1-\beta) (x_1Y + X y_1-X Y) \nonumber\\
	& & +\beta y_1(t-\tau) Z -b Z, \nonumber 
\end{eqnarray}
where $X=x_1-x_2$, $Y=y_1-y_2$, and $Z=z_1-z_2$.
The above equations are nonautonomous in themselves and, so,
by iterating the above equation with Eqs. (3)-(4), we obtain the
transverse Lyapunov exponents (TLEs) \cite{TLE} which enable us 
to determine the synchronization threshold.

\begin{figure*}
\begin{center}
\rotatebox[origin=c]{0}{\includegraphics[width=14.0cm]{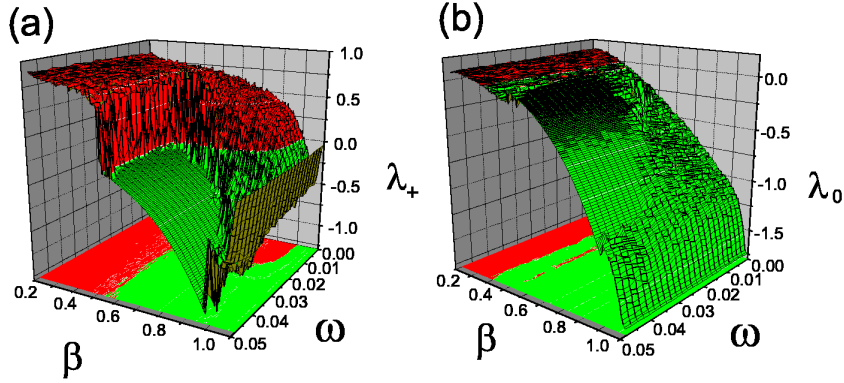}}
\caption{Two largest TLEs as a function of
	the coupling $\beta$ and the modulation frequency $\omega$. 
	We have added the uniform random noise of amplitude $ |\xi| \sim 10^{-8}$ only to 
	$x$ variable of Eq. (3) 
	in order to remove abrupt synchronization regime:
	(a) the largest exponent, where the border line between gray and dark gray region on $(\beta, \omega)$ 
	plane corresponds to 
	the transition threshold to CS; (b) the second largest exponent, 
	where the border line corresponds to the transition threshold to PS.}
\end{center}
\end{figure*}

\begin{figure}
\begin{center}
\rotatebox[origin=c]{0}{\includegraphics[width=7.3cm]{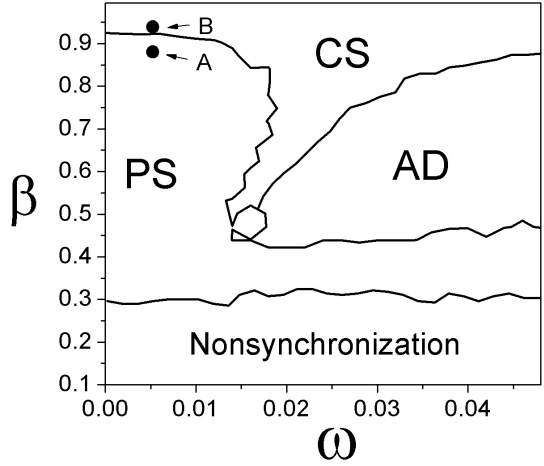}}
\caption{Contour plot of synchronization regimes in $(\omega, \beta)$ space.
	A and B are the reference points to show the time series.
	CS: Complete Synchronization, PS: Phase Synchronization, AD: Amplitude Death.}
\end{center}
\end{figure}

Figure 3 shows two largest TLEs as a function of $\beta$ and $\omega$.
One can clearly see the threshold to CS on $(\beta, \omega)$ plane of Fig. 3 (a)
where the largest TLE becomes negative. 
Also the threshold to PS, where the vanishing TLE  becomes negative,
is plotted on  $(\beta, \omega)$ plane of Fig. 3 (b).  
The synchronization regimes are presented in Fig. 4 as a contour plot.
Contrary to a common intuition, the CS regime becomes wider as the
modulation frequency $\omega$ increases, even though the regime
is smeared with an amplitude death (AD) \cite{AD,Death}. 
The amplitude death phenomenon due to DTM  in our study is similar to 
the transition to periodic orbit in Ref. \cite{Madruga} in the sence that
in both cases the systems lose their exponential instability by the interaction. 
However different apect is that our system converges to the fixed point 
without oscillation instead of the periodic orbit.
These effects show that DTM can play a role of stabilizing scheme such as chaos control\cite{TD_Cont}.
The appearance of the wider synchronization regime in Fig. 4 implies that the scheme  
will be effective to implement real communication systems, 
since the higher modulation frequency widens the synchronization regime
and it would enable us to construct more of communication channels.  
\begin{figure}
\begin{center}
\rotatebox[origin=c]{0}{\includegraphics[width=8.3cm]{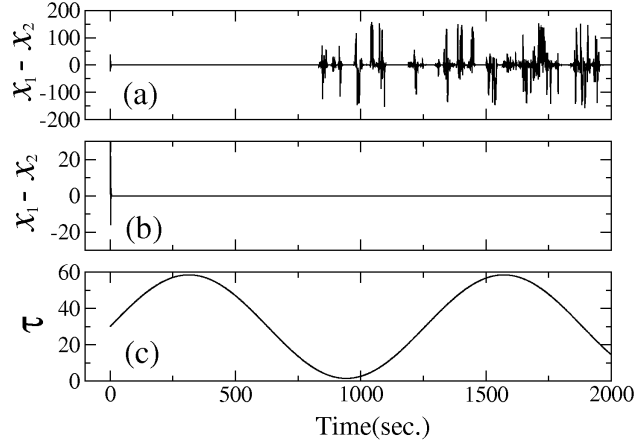}}
\caption{Time series of $x_1-x_2$ and the delay time $\tau$ as a function of time:  
	(a) point A in Fig. 4 ($\beta=0.87, \omega=0.005$). 
	(c) point B in Fig. 4 ($\beta=0.93, \omega=0.005$). 
	(c) modulated delay time when $\omega=0.005$.}
\end{center}
\end{figure}

Figure 5 shows the temporal behaviors of the difference between $x_1$ and $x_2$
and the modulated delay time $\tau$.
The temporal behaviors below (point A in Fig. 4) and above (point B in Fig. 4) 
the threshold are presented in Fig. 5 (a) and Fig. 5 (b), respectively.
They clearly show the threshold behaviors of CS at two points.  
The delay time $\tau$ is modulated as shown in Fig. 5 (c). 
It is interesting to see the autocorrelations and filling factor for different modulations.
For the constant delay time, 
the delay time is exactly indicated by the position of
the dips pointed by arrows in Fig. 6 (a) and (c).
In the case of DTM (Fig. 6 (b) and (d)), 
the dips disappear.

In general introducing a $d$-dimensional chaotic oscillator,
we can consider a chaotic modulation of delay time:
$\tau = g({\bf x}, \tau_0)$ 
where
$\dot{\bf x} = \frac{1}{\epsilon}{\bf F}({\bf x})$ and  
$\epsilon$ is a time scaling parameter to control
the rate of modulation like $\omega$ in sinusoidal modulation.
We can expect that the chaotic modulation
will not work a fundamental change in contrast with the sinusoidal modulation, 
because the delayed feedback $x(t-\tau)$ hardly correlated with $x(t)$ already 
even in the sinusoidal modulation as shown in Fig. 6 (b).
However, in the application to communication, the chaotic modulation 
would enhance the security of the system 
because it may increase the complexity of the attractor. 
\begin{figure}
\begin{center}
\rotatebox[origin=c]{0}{\includegraphics[width=8.0cm]{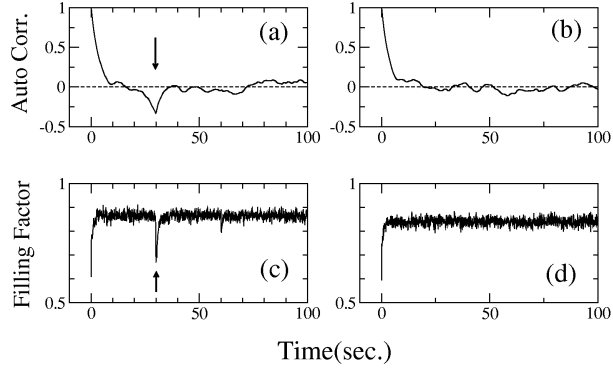}}
\caption{Disappearance of the dips in autocorrelations and filling factors at $\beta=0.93$:
	(a) and (c) constant delay time, $\tau_0=30$; (b) and (d) modulated delay time ($\omega=0.005$).}
\end{center}
\end{figure}
It is worth discussing the physical implementation of the systems described by
Eq. (3)-(4).  We think that such a situation can be observed in optical systems.
For example, one can consider two laser systems coupled by reflected laser beam from a vibrating mirror.
In that case, Lorenz equation describes a laser system and 
the vibrating mirror plays a role of modulation function.

In conclusion, we have studied the effects of  DTM in chaotic oscillators.  
We have observed that the imprints due to the constant time delay 
are completely wiped out by the modulation in the chaotic map and flow. 
We have also demonstrated that the robust synchronization can be established 
in the presence of DTM.  
We expect this effects to be useful in constructing real communication systems.

We thank M. S. Kurdoglyan for helpful discussions.
This work is supported by Creative Research Initiatives 
of the Korean Ministry of Science and Technology.


\begin{thebibliography}{150}
\bibitem{Huy} Ch. Huygens, {\it Horologium Oscillatiorium} Apud F. Muguet, Paris, France, 1673 English
		translation: {\it The Pendulum Clock} Iowa State University Press, Ames, 1986.
\bibitem{SyncOrg0} H. Fujisaka and T. Yamada, Prog. Theor. Phys. {\bf 69}, 32 (1983);
		   V.S. Afraimovich, N.N. Verichev, and M.I. Rabinovich, Radiophys. 
		   Quantum Electron. {\bf 29}, 747 (1986);
		    L.M. Pecora and T.L. Carroll, Phys. Rev. Lett. {\bf 64}, 821 (1990).
\bibitem{SyncTut} S. Boccaletti, J. Kurth, G. Osipov, D.L. Valladares and C. Zhou, Physics Reports {\bf 366}, 1 (2002) and references therein. 
\bibitem{SyncBook} A. Pikovsky, M. Rosenblum, and J. Kurths, {\it Synchronization A universal concept in nonlinear
			science},  CAMBRIDGE UNIVERSITY PRESS, 2001. 
\bibitem{GPS} D.-S. Lee, W.-H. Kye, S. Rim, T.-Y. Kwon, and C.-M. Kim,
		 Phys. Rev. E{\bf 67}, 045201(R)(2003). 
\bibitem{NLDF}K.M. Short and A.T. Parker, Phys. Rev. E {\bf 58}, 1159 (1998) and references therein. 
\bibitem{TD_Sync} K. Pyragas, Phys. Rev. E {\bf 58}, 3067 (1998);
		 R. He and P.G. Vaidya, Phys. Rev. E {\bf 59}, 4048 (1999);
		 L. Yaowen, G, Gguangming, Z. Hong, and W. Yinghai, Phys. Rev. E {\bf 62}, 7898 (2000).
\bibitem{TD_Laser} T. Heil, I. Fischer, W. Els\"asser, J. Mulet, and C. R. Mirasso, Phys. Rev. Lett. {\bf 86}, 2001 (795).
\bibitem{TD_Circuit} D. V. Ramana Reddy, A. Sen, and G.L. Johnston, Phys. Rev. Lett. {\bf 85}, 3381 (2000).
\bibitem{TD_Cont} K. Pyragass, Phys. Rev. Lett. {\bf 86}, 2265 (2001);
		O. L\"uthje, S. Wolf, and G. Pfister, Phys. Rev. Lett. {\bf 86}, 1745 (2001).
\bibitem{TD_Comm} V.S. Udaltsov, J.-P. Goedgebuer, L. Larger, and W.T. Rhodes, Phys. Rev. Lett. {\bf 86}, 1892 (2001).
\bibitem{Hyperchaos} J.D. Farmer, Physica D {\bf 4}, 366 (1982).
		    K.M. Short and A.T. Parker, Phys. Rev. E {\bf 58}, 1159 (1998).
\bibitem{AutoCorr} F.T. Arecchi, R. Meucci, E. Allaria, A. Di Garbo, and L.S. Tsimring, Phys. Rev. E{\bf 65}, 046237 (2002).
\bibitem{Bunner} M.J. B\"unner, Th. Meyer, A. Kittel, and J. Parisi, Phys. Rev. E {\bf 56}, 5083 (1997); 
		 R. Hegger, M.J. B\"unner, H. Kantz, and A. Giaquinta, Phys. Rev. Lett. {\bf 81}, 558 (1998);
		M.J. Buenner, M. Ciofini, A. Giaquinta, R. Hegger, H. Kantz, R. Meucci and A. Politi, 
		Eur. Phys. J. D {\bf 10}, 165 (2000);
		M.J. Buenner, M. Ciofini, A. Giaquinta, R. Hegger, H. Kantz, R. Meucci and A. Politi, 
		Eur. Phys. J. D {\bf 10}, 177 (2000); {\it ibid} {\bf 10}, 177 (2000).
\bibitem{TD_ATT} V.I. Ponomarenko and M.D. Prokhorov, Phys. Rev. E {\bf 66}, 026215 (2002);
		 C. Zhou and C.-H. Lai, Phys. Rev. E {\bf 60}, 320 (1999);
		 B.P. Bezruchko, A.S. Karavaev, V.I. Ponomarenko, and M.D. Prokhorov,
		 Phys. Rev. E {\bf 64}, 056216 (2001).
\bibitem{Madruga} S. Madruga and S. Boccaletti, and M.A. Matias, International Journal of Bifurcation
		and Chaos {\bf 11}, 2875 (2001). 
\bibitem{FF} 
	Consider the $({ \bf x}(t), {\bf x}(t-\tau), \dot{\bf x})$ space 
	with $P^{3N}$ equally sized hypercubes.
	The filling factor is the number of hypercubes, which are visited by
	the projected trajectory, normalized to the total number of hypercubes, $P^{3N}$ (see the
	first reference in Ref \cite{Bunner}).
	We have used $(x(t), x(t-\tau), \dot{x})$ space for filling factors in Fig. 1 (c) and (d)
	and $(x_1, y_1(t-\tau), \dot{x}_1)$  space for filling factors in Fig. 6 (c) and (d).
\bibitem{TLE} J.F. Heagy, T.L. Carroll, and L.M. Pecora, Phys. Rev. E{\bf 50}, 1874(1994).
\bibitem{AD}
In the amplitude death regime, the amplitudes of oscillation of state vectors ${\bf x}_1$ and ${\bf x}_2$ 
are strongly damped and eventually converge to fixed points.
Thus, in this regime, the transverse variables $X$, $Y$, and $Z$ converges to zero.
This is the first observation of amplitude death due to DTM in chaotic oscillator, as far as we know. 
We will report the phenomenon with detail analyses elsewhere; for understanding of amplitude
death in coupled limit cycles, see, for example, Ref. \cite{Death}. 
\bibitem{Death} D.V. Ramana Reddy, A. Sen, and G.L. Johnston, Phys. Rev. Lett. {\bf 80}, 5109 (1998).
\end{thebibliography}
\end{document}